\documentclass[namedreferences]{solarphysics}

\usepackage[hyperref,optionalrh,showbiblabels]{spr-sola-addons} 
\usepackage{graphicx}
\usepackage{threeparttable}
\usepackage{ulem}
\usepackage{amssymb}        
\usepackage{color}           
\usepackage{breakurl}        

\renewcommand{\vec}[1]{{\mathbfit #1}}

\chardef\us=`\_

\begin{document}

\begin{article}
\begin{opening}

\title{Evaluating Pointing Strategies for Future Solar Flare Missions}

\author[addressref = {aff1,aff2}]{Andrew R. Inglis}
\author[addressref = aff1]{Jack Ireland}
\author[addressref = aff1]{Albert Y. Shih}
\author[addressref = aff1]{Steven D. Christe}

\address[id=aff1]{Solar Physics Laboratory, NASA Goddard Spaceflight Center, Greenbelt, MD 20771, USA.}
\address[id=aff2]{Physics Department, The Catholic University of America, Washington, DC, 20064}

\runningauthor{A. R. Inglis et al.}
\runningtitle{Pointing Strategies for Solar Flare Missions}

\begin{abstract}
Solar flares are events of intense scientific interest. Although certain solar conditions are known to be associated with flare activity, the exact location and timing of an individual flare on the Sun cannot as yet be predicted with certainty. Missions whose science objectives depend on observing solar flares must often make difficult decisions on where to target their observations if they do not observe the full solar disk. Yet, little analysis exists in the literature which might guide these missions' operations to maximize their opportunities to observe flares. In this study we analyze and simulate the performance of different observation strategies using historical flare and active region data from 2011 to 2014. We test a number of different target selection strategies based on active region complexity and recent flare activity, each of which is examined under a range of operational assumptions. In each case we investigate various metrics such as the number of flares observed, the size of flares observed, and operational considerations such as the number of instrument re-points that are required. Overall, target selection methods based on recent flare activity showed the best overall performance, but required more repointings than other methods. The mission responsiveness to new information is identified as another strong factor determining flare observation performance. It is also shown that target selection methods based on active region complexities show a significant pointing bias towards the western solar hemisphere. As expected, the number of flares observed grows quickly with field-of-view size until the approximate size of an active region is reached, but further improvements beyond the active region size are much more incremental. These results provide valuable performance estimates for a future mission focused on solar flares, and inform the requirements that would ensure mission success.

\end{abstract}

\end{opening}

\section{Introduction}
\label{introduction}

Solar flares are among the most energetic phenomena in the solar system, releasing up to 10$^{32}$ ergs in a matter of minutes. In combination with coronal mass ejections (CMEs), they are a key driver of space weather. Space-based observations of flares have been carried out for over 50 years, with varying degrees of success in observing solar flares. The largest challenge facing any solar flare mission is predicting when and where a flare will occur, and flare prediction remains a topic of active research. 

Several recent heliophysics missions have observed the Sun with partial Sun field-of-view (FOV) instruments, including the Transition Region and Coronal Explorer \citep[TRACE;][]{1999SoPh..187..229H}, the Coronal Diagnostic Spectrometer \citep[CDS;][]{1995SoPh..162..233H} and the Solar Ultraviolet Measurements of Emitted Radiation \citep[SUMER;][]{1995SoPh..162..189W} instruments on board the Solar and Heliospheric Observatory \citep[SOHO;][]{1995SoPh..162....1D}, the Interface Region Imaging Spectrograph \citep[IRIS;][]{2014SoPh..289.2733D}, the Solar Optical Telescope \citep[SOT;][]{2008SoPh..249..167T} and the EUV Imaging Spectrometer \citep[EIS;][]{2007SoPh..243...19C} on board the Hinode mission \citep{2007SoPh..243....3K}, as well as the High Resolution Coronal Imager \citep[Hi-C;][]{2014SoPh..289.4393K, 2019SoPh..294..174R}, and the Focusing Optics X-ray Solar Imager \citep[FOXSI;][]{2009SPIE.7437E..05K, 2014ApJ...793L..32K, 2016JAI.....540005C} sounding rocket missions. These instruments have had to decide where on the Sun to observe in order to maximize the probability of observing solar flares. Each team has relied on different approaches with different results. These approaches have generally not been published or analytically evaluated.

In this paper we investigate the performance of a space-based solar flare mission using different operational strategies to maximize the number of flares observed using historical solar flare and active region data. We explore the fraction of available solar flares a mission might expect to observe under a range of science operations scenarios, which solar targeting strategies are most effective, whether there are intrinsic biases associated with different strategies, as well as the impact of mission responsiveness to new information and the impact of the size of the field of view. In Section \ref{data} we describe the solar flare and active region data sources used throughout this paper. In Section \ref{simulation} we discuss the mission operations simulation and the various strategies and scenarios used to observe solar flares. In Section \ref{results} we present the detailed results of this work. The key findings are summarized in Section \ref{discussion}.

\section{Data}
\label{data}

\begin{figure}
    \centering
    \includegraphics[width=12cm]{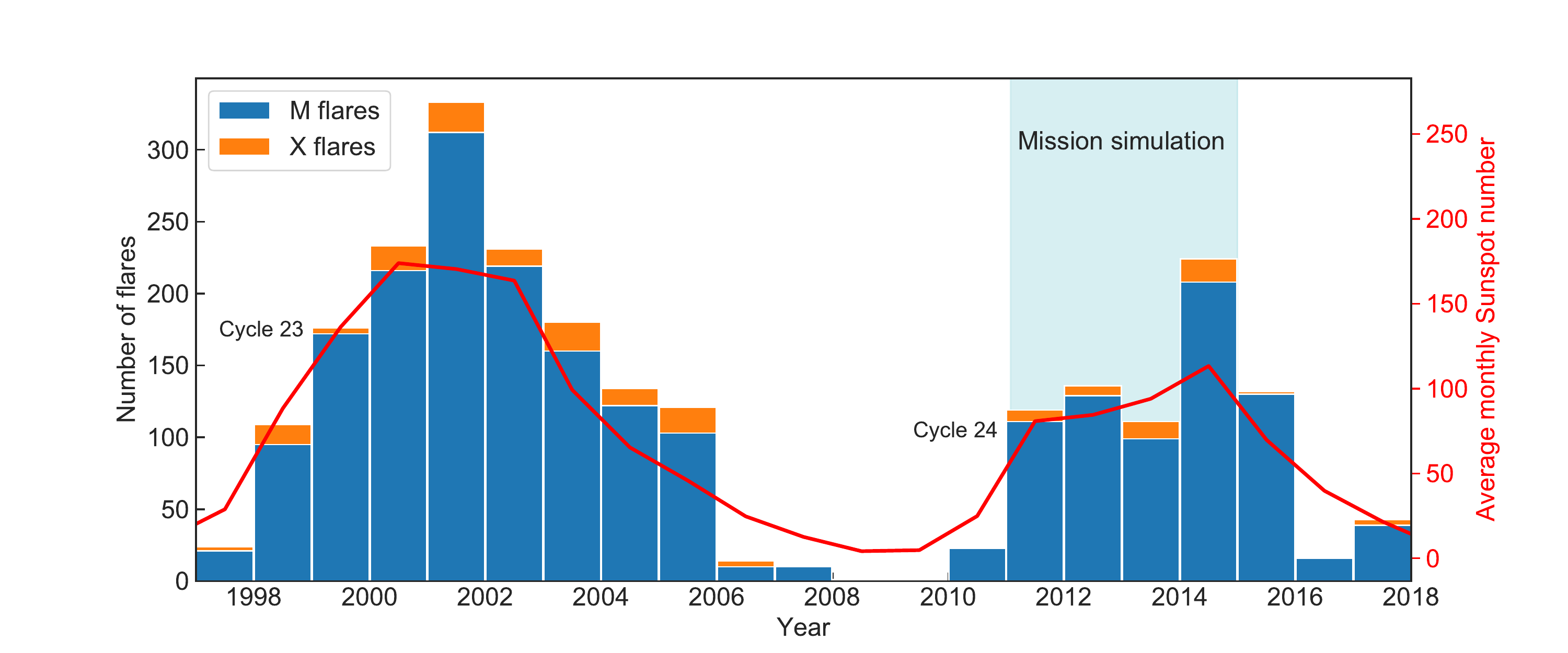}
    \caption{Flare activity during Solar Cycles 23 and 24. The number of M-class (blue) and X-class (orange) flares occurring each year according to the GOES event list is shown, along with the year-averaged monthly sunspot number (red), available from the Solar Influences Data analysis Center (SIDC). The time period of our mission simulation is 2011 February 1 to 2014 December 31 and is represented by the blue shaded area.}
    \label{solar_cycle}
\end{figure}

To better understand the flare observing performance of a partial field-of-view solar flare mission, we choose to simulate its operation using readily available data sources which are consistent and mostly free of gaps in observations and known errors. For this reason we choose an observational time window within the Solar Dynamics Observatory (SDO) era. Specifically, the mission observation period in this study is defined as 2011 February 1 until 2014 December 31. This period corresponds to the rise and maximum phase of Solar Cycle 24, thus providing a relevant reference time frame for a future solar flare mission (see Figure \ref{solar_cycle}). Data sources in this time period are relatively consistent and need a minimum of further cleaning to make them useful for our simulations.  The principal data sources required are accurate records of all the solar flares that occurred, as well as a comprehensive list of active regions present on the Sun during that time. 

\subsection{Active Region Information}
\label{ss:arinfo}
The National Oceanic and Atmospheric Administration (NOAA) provides daily information on all visible active regions on the solar disk through the Space Weather Prediction Center (SWPC), typically published at 00:30 UTC. This product is known as the Solar Region Summary (SRS). This product contains the key properties of active regions required for this study: NOAA active region number, location, McIntosh and Hale classifications, as well as other useful parameters such as active region area and number of associated sunspots. 

\subsection{Solar Flare Information}
\label{ss:flareinfo}
A consistent and long-running database of flares is produced by NOAA based on Geostationary Operational Environment Satellite (GOES) X-ray Sensor (XRS) soft X-ray data, which we refer to as the GOES event list. The source of the GOES event list information used in this paper was event summary files produced by the Space Weather Prediction Center (SWPC). These were obtained by querying the Heliophysics Event Knowledgebase (HEK) for records of solar flares using appropriate keywords. One advantage of this product is that continuous GOES observations of the Sun have been available for decades, providing an uninterrupted record of flare properties with a consistent methodology. These records also provide the well-known GOES classifications (e.g. `M2', `X1') for each flare, their locations on the Sun, as well as their source active region designations (e.g. `AR 11158'). The records also list the start, peak, and end times for each flare. The GOES end time is defined as the point where the flare's
excess flux above the pre-flare background has receded to 50\% of the peak excess flux.

In this work we are focused on large flares, which we define as GOES class M1 or greater. During the period from 2011 February 01 to 2014 December 31, there were 589 M- and X-class flares reported in the GOES event list (see Figure \ref{solar_cycle}). Of these, 259 entries are recorded with a heliographic position of (0,0), indicating no information on position in the GOES event list. We determine a position using a two-step process. First, we cross-check each flare in the GOES event list with the Solar Flare Finder (SFF) database \citep{2018SoPh..293...18M}, which provides flare location information for the majority of the 2010 -- 2017 time range. For every GOES flare list entry with a missing position, we extract and use the position recorded for the matching flare in the SFF database. Following this correction, there are 69 events remaining with no position information. This is primarily due to gaps in coverage of the SFF database \citep[see][for details]{2018SoPh..293...18M}. For these remaining flares, we manually correct their positions using SDO/AIA data, first using the coarse flare position identified by the SDO Flare Detective algorithm \citep{2010AAS...21640208G}, followed by visual verification of the position and further correction if needed.

\section{Mission Simulation}
\label{simulation}

Using the solar active region and flare data described in Section \ref{data} as inputs, we can simulate difference science operations approaches to maximize flare observations. From this, we can assess the likely performance of a flare mission in a range of scenarios, and identify the key factors that affect that performance. As described in Section \ref{data}, the time period of our mission simulation begins on 2011 February 1 and ends on 2014 December 31. We simulate a mission in low-earth orbit (LEO) placed in a medium-inclination configuration, in this case 600 km at 28.5$^{\circ}$. Thus, the mission is subject to eclipse intervals and passes through the South Atlantic Anomaly (SAA). The total time spent in eclipse in this configuration is 35\%. Meanwhile, 9\% of the total time is spent in the SAA, where we assume solar observations cannot be made. Since SAA and eclipse intervals sometimes overlap, the total time unable to observe the Sun is 41\%, leaving a solar observing duty cycle of 59\%. Unless otherwise stated the field-of view of the instrument is 9.8 $\times$ 9.8 arcminutes, corresponding to that of the previously proposed FOXSI Small Explorer (SMEX) mission \citep{2019AAS...23422501C}.

Since the SRS summary product is published daily, and science missions often plan on a daily schedule, we adopt this cadence for our simulations of science operations (exceptions to this daily schedule are discussed in Section \ref{flare_numbers}).

On each day during the mission, we use the published active region and solar flare data from the previous day as inputs to select where on the Sun to point to. Since large flares typically only occur in active regions, we choose the pointing target from among the currently numbered and visible NOAA active regions on the solar disk. Once the target is determined, the mission FOV is centered on the target and interpolation is used to follow the target as it moves across the solar disk, accounting for solar rotation. This continues until a new target is chosen. The process for choosing a new target begins at 00:30 UT, corresponding to the time that new SRS reports are released. We consider several different target determination methods, which are described in detail in Section \ref{target_methods}. In realistic science operations there is often a considerable delay between new information becoming available and the choice of a new mission target and subsequent commanding of space-based assets. First, there is a delay due to science team deliberations and the work schedule of the science team. Following this, additional time is needed  to construct and verify new spacecraft commands, to wait for an available communications pass, and to accommodate the work schedule of the operations team. To reflect these factors, the pointing change in our simulations takes place after a time delay $t_r$ from 00:30 UT. We test several distributions for $t_r$, which are discussed in Section \ref{flare_numbers}. To determine which M- and X-class solar flares were observed during the mission, we compare the mission pointing over time each day with the locations of flares that occurred and the eclipse and SAA interval data. For this purpose, a solar flare is considered observed if any portion of it between the listed GOES start and end times is observable (i.e. not in eclipse or SAA) and in the field-of-view.

To evaluate a given scenario, we use a Monte Carlo method, running many iterations of a simulation with randomized elements to build up a distribution of results. The flare and active region data itself is fixed since it is historical data. The time delay $t_r$ between choosing a target with new SRS data and the commanding of the spacecraft is defined by a normal distribution, with the delay for each individual pointing change drawn randomly from that distribution. Most operations scenarios also include eclipse and SAA intervals. The timing of these intervals is determined from the mission orbital profile. For each iteration of the simulation, the mission start time is randomized with a shift of up to 24 hours, thus adjusting the timing of the eclipse and SAA intervals for that iteration of the simulation. This removes any systematic effects that could arise from flare times consistently coinciding with eclipse intervals. For each scenario we perform $N$ = 1000 iterations of the simulation.

\subsection{Methods for Determining Pointing Target}
\label{target_methods}

Flare prediction is an active and developing field of research \citep[e.g.][]{1994SoPh..150..127B, 2002SoPh..209..171G, 2008ApJ...688L.107B, 2012ApJ...747L..41B, 2013SoPh..283..157A, Falconer, 2015ApJ...798..135B, 2016ApJ...829...89B, 2017SpWea..15..577M, 2018JSWSC...8A..34M, 2018SoPh..293...96K, 2019ApJ...883..150C, 2019ApJS..243...36L, 2019ApJ...881..101L}. The goal of our work is not to further the field of flare prediction, but to investigate the potential performance of current and future flare-focused missions given target selection strategies and operational constraints. This problem simplifies to pointing to the active region most likely to produce a flare. Since we are using the SRS data product, the target is always a valid numbered NOAA active region unless there are no such regions visible on the solar disk, in which case the default target is the center of the solar disk. The targeting methods are chosen to be straightforward and to make use of currently available active region and flare summary data. In the following sections we describe the approaches and their motivation.

\subsubsection{Target Based on Hale Classification Plus Number of Sunspots}

One system for describing the complexity of solar active regions is the Hale (or Mt Wilson) classification, first introduced a century ago by \citet{1919ApJ....49..153H} and subsequently modified by \citet{1960AN....285..271K}. The Hale classification provides a simpler system for categorizing the complexity of active regions compared with the McIntosh system (see Section \ref{mcintosh_target_description}).

A number of studies have investigated the number of flares produced by each Hale active region class \citep{2000ApJ...540..583S, 2014MNRAS.441.2208G}. The majority of strong flares are sourced from $\beta \gamma \delta$-type regions, with the $\delta$ designation being particularly important for flare productivity. However, these studies did not explicitly include the occurrence rates of each different Hale class. Recent work by \citet{2016ApJ...820L..11J} investigated the occurrence distribution of active region Hale classifications between 1992 -- 2015. By combining the active region Hale class distribution provided by \citet{2016ApJ...820L..11J} with the flare occurrence data in \citet{2014MNRAS.441.2208G} we construct an estimate of the flare productivity of each Hale class, and generate a rank order for target selection.

However, given the relatively small number of possible Hale classifications, it is common for multiple active regions of equivalent type to co-exist, i.e. there may be multiple $\beta \gamma \delta$ regions present simultaneously. Therefore, additional information is required in order to select a single target. To break ties between regions with the same Hale classification, we select the active region with the largest number of sunspots.

\subsubsection{Target Based on McIntosh Classification}
\label{mcintosh_target_description}

The McIntosh classification system has been used to provide a measure of active region complexity for decades \citep{1990SoPh..125..251M}. This system provides a three-component description of an active region, with 60 possible classifications in total. The first parameter represents the modified Zurich class, which defines whether a penumbra is present, the distribution of that penumbra and the length of the sunspot group. The second parameter describes the type of penumbra surrounding the largest spot of the group, while the third parameter describes the degree of sunspot compactness in the group. It has been shown that more complex active regions are more flare-productive. In particular, \citet{1994SoPh..150..127B} showed the correlation between X-ray flare activity and each possible Mcintosh classification. This data was later combined with more recent observations by \citet{2012ApJ...747L..41B}, providing more up-to-date flare productivity estimates. From this analysis, the most flare-productive regions (for larger flares) were complex types such as $Ekc$, $Fkc$, $Ehc$, $Fsi$ and $Fki$, while simple active regions types such as $Axx$ and $Bxo$ occur frequently but produce few flares (see Appendix \ref{mcintosh_appendix}). Other studies carried out on recent datasets produced results consistent with this picture \citep[e.g.][]{2006AN....327...36T, 2011SoPh..269..111N}, although the precise rank-ordering of McIntosh classifications varies between studies.

Thus, the McIntosh classification can be used to estimate the flare productivity of an active region. In this work, we adopt the flare productivity values published by \citet{2012ApJ...747L..41B} in order to rank available solar active regions. Alternative estimates of the flare productivity from Mcintosh classifications are discussed in Appendix \ref{mcintosh_appendix}.

\subsubsection{Target Based on Prior Flaring Activity}
 
One of the most reliable predictors of future flare activity is that flares have already occurred in an active region \citep[e.g.][]{2019ApJ...883..150C}. This method alone can never be 100\% predictive, since an observer must by definition wait for some flares to occur in an active region before being able to choose it as a target.

We implement this strategy by defining the flare index for each active region as the sum of the GOES X-ray peak fluxes from each flare that occurred in the previous 24 hours \citep{2019ApJ...883..150C}. The active region with the largest flare index is selected as the target. This means that a newly identified active region will never be pointed to in the first 24 hours -- unless no other targets are available -- since its flare index will be undefined. In order to have sufficient information to construct a flare index, we include C-class flares in this calculation. This is necessary as even during solar maximum, there are many days without M or X-class flares (see also Section \ref{ar_exploration}).

\subsubsection{Random Target}

 It is also useful to compare the above methods with an intentionally information-starved methodology, in order to provide a point of comparison. Hence, we also implement a purely random target selection. In this setup, each day an arbitrary active region target is chosen from the list of available regions. This represents a lower limit or worst case for comparison. The performance of each data-based strategy must demonstrate superior performance compared with this method.

\section{Results}
\label{results}

\subsection{Flare Observing Performance of Different Strategies}
\label{flare_numbers}

We first investigate the nominal performance of each chosen flare pointing strategy via our Monte Carlo mission simulation. As discussed in Section \ref{data}, each day new information regarding solar active regions is made available at approximately 00:30 UT. At this time the process of choosing the most promising active region target begins, using strategies defined in Section \ref{target_methods}. The delay between 00:30 UT and the mission pointing to the new target is given by $t_r(\mu, \sigma, t_{min})$, a randomized response time drawn from a normal distribution with mean $\mu$, standard deviation $\sigma$ and a minimum allowed value $t_{min}$. When values are drawn from the distribution that would be smaller than $t_{min}$, they are replaced by $t_{min}$. The quantities $\mu$, $\sigma$ and $t_{min}$ are given in units of hours.

We examine six distinct delay scenarios (see Table \ref{results_table}). For the first scenario, we simulate a mission with $t_{r}(24, 2, 20)$ that only responds to new information once per week, while the second scenario responds three times per week, also with a delay of  $t_{r}(24, 2, 20)$.  For the next three scenarios, we simulate a mission with daily response capability to new information, with delays of $t_{r}(24, 2, 20)$, $t_{r}(12, 2, 10)$  and $t_{r}(1, 0.5, 0)$ respectively. We informally refer to these $t_r$ values as the `24 hour delay', `12 hour delay', and `1 hour delay'. For the sixth scenario, we implement a response time $t_{r}(1, 0.5, 0.0)$, but this time for a mission that does not experience either eclipses or SAA intervals, equivalent to a mission in geostationary orbit (e.g. GOES) or at Lagrange point $L_1$ (e.g. SOHO). These scenarios are summarized in Table \ref{results_table}; for convenience each scenario is given an identifier, e.g. `M-4' for the McIntosh target method with daily response capability and a 12-hour repoint delay.

\begin{figure}
    \centering
    \includegraphics[width=10cm]{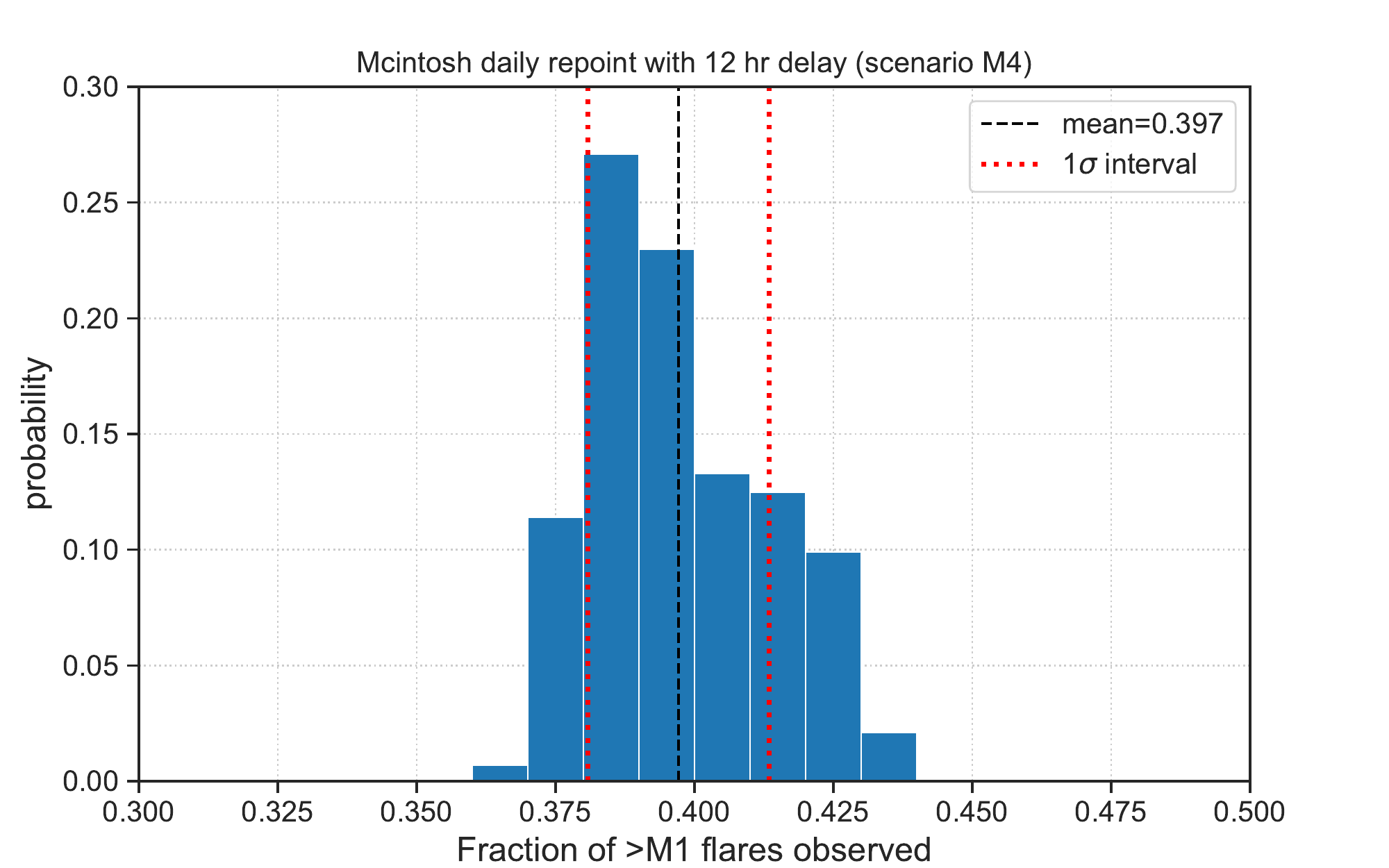}
    \caption{Distribution of the fraction of available $>$M1 solar flares observed using scenario M-4, the McIntosh-based targeting strategy with a mean delay time of 12 hours ($t_r(12,2,10)$). The mean of the distribution is 0.397, with minimum and maximum values of approximately 0.36 and 0.44 respectively.}
    \label{performance_example}
\end{figure}

\begin{figure}
    \centering
    \includegraphics[width=12cm]{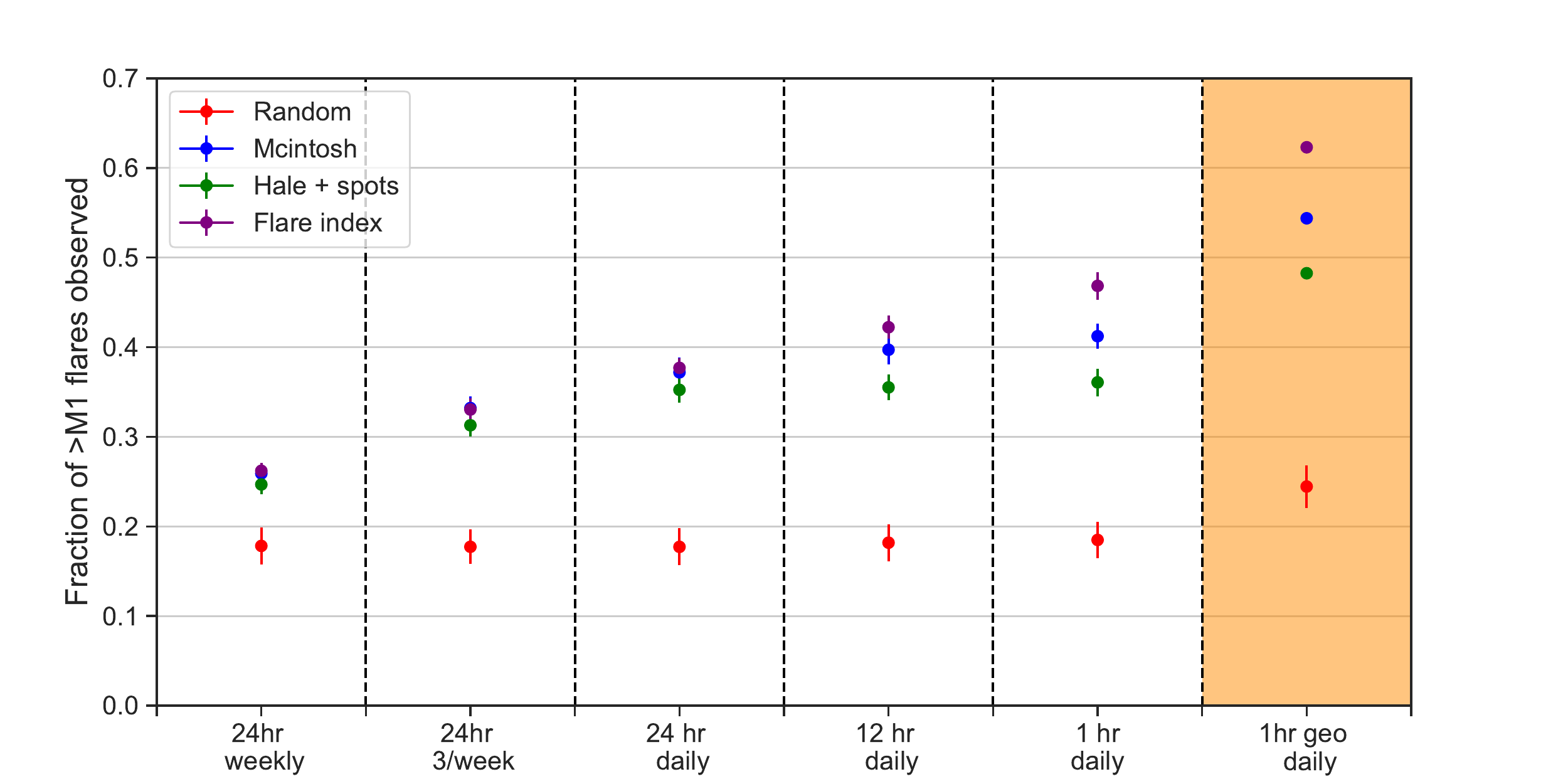}
    \caption{The performance of different observing strategies and response times in terms of the number of flares successfully observed. The mean fraction of $>$M1 solar flares observed is shown for each combination of target method and response time. The Mcintosh-based target results are shown in blue, the Hale-based results in green, the flare index results in purple, and the random target results in red. Each result is from a simulation with $N$=1000 iterations.}
    \label{overall_performance}
\end{figure}

An example distribution of the number of flares observed is shown in Figure \ref{performance_example}, featuring the McIntosh-based target method and a daily response with a mean delay of 12 hours (scenario M-4). On average, around 40\% of available flares are observed, with the full distribution ranging from a low of 36\% to a high of 43\%. Figure \ref{overall_performance} shows the fraction of M and X class flares observed as a result of each strategy and delay scenario combination. This information is also summarized in Table \ref{results_table}. In this context a flare is considered observed if it is in the field of view and any part of the flare is observable between the flare start and end times defined in the GOES catalogue, i.e. a flare fully obscured by an eclipse or SAA interval will not be counted. Overall, the most effective target method for observing solar flares among those tested is the flare index, slightly outperforming target selection via the McIntosh classification at all response times except for the weekly and thrice-weekly repoint scenarios, where the performance is equal. Past flare activity has been considered a predictor of future flares for many years, a finding that has been replicated by recent machine learning studies \citep{2019ApJ...883..150C}. However, it is notable that this index alone matches and even outperforms approaches that directly describe active region complexity such as the McIntosh and Hale classifications. 

\begin{table*}[]
\centering
\caption{Summary of the performance of different observing strategies, decision frequencies, and re-point delay times.}
\label{results_table}
\begin{tabular}{c|ccccc}
\hline
Scenario & Targeting & Decision & Repoint$\dagger$ & Mean \% & Mean time \\
ID & strategy & frequency & delay & of flares obs. & between \\
& & & & $\pm$ (1$\sigma$) & repoints (days) \\
\hline
H-1 & Hale & Weekly & 24hr & 24.7 $\pm$ 1.1 & 7.6 \\
H-2 & Hale & 3/week & 24hr & 31.3 $\pm$ 1.3 & 4.1 \\
H-3 & Hale & Daily & 24hr & 35.2 $\pm$ 1.4 & 2.8\\
H-4 & Hale & Daily & 12hr & 35.5 $\pm$ 1.4 & 2.8 \\
H-5 & Hale & Daily & 1hr & 36.1 $\pm$ 1.5 & 2.8\\
H-6 & Hale (geo) & Daily & 1hr & 48.3 $\pm$ 0.2 & 2.8 \\
\hline
M-1 & McIntosh & Weekly & 24hr & 25.9 $\pm$ 0.9 & 7.8 \\
M-2 & McIntosh & 3/week & 24hr & 33.2 $\pm$ 1.3 & 4.5 \\
M-3 & McIntosh & Daily & 24hr & 37.2 $\pm$ 1.6 & 2.6 \\
M-4 & McIntosh & Daily & 12hr & 39.7 $\pm$ 1.6 & 2.6 \\
M-5 & McIntosh & Daily & 1hr & 41.2 $\pm$ 1.4 & 2.6 \\
M-6 & McIntosh (geo) & Daily & 1hr & 54.4 $\pm$ 0.2 & 2.6 \\
\hline

F-1 & Flare index & Weekly & 24hr & 26.2 $\pm$ 0.9 & 8.0 \\
F-2 & Flare index & 3/week & 24hr & 33.0 $\pm$ 1.2 & 3.4 \\
F-3 & Flare index & Daily & 24hr & 37.7 $\pm$ 1.1 & 1.9\\
F-4 & Flare index & Daily & 12hr & 42.2 $\pm$ 1.3 & 1.9\\
F-5 & Flare index & Daily & 1hr & 46.8 $\pm$ 1.5 & 1.9\\
F-6 & Flare index (geo) & Daily & 1hr & 62.3 $\pm$ 0.3 & 1.9\\
\hline
R-1 & Random & Weekly & 24hr & 17.8 $\pm$ 2.0 & 7.4  \\
R-2 & Random & 3/week & 24hr & 17.7 $\pm$ 1.9 & 2.7 \\
R-3 & Random & Daily & 24hr & 17.7 $\pm$ 2.1 & 1.2\\
R-4 & Random & Daily & 12hr & 18.2 $\pm$ 2.1 & 1.2 \\
R-5 & Random & Daily & 1hr & 18.5 $\pm$ 2.0 & 1.2\\
R-6 & Random (geo) & Daily & 1hr & 24.4 $\pm$ 2.4 & 1.2 \\
\hline

\end{tabular}
\begin{tablenotes}
\item $\dagger$ `24hr' delay corresponds to $t_r(24,2,20)$
\item `12hr' delay corresponds to $t_r(12,2,10)$ 
\item `1hr' delay corresponds to $t_r(1,0.5,0)$.
\end{tablenotes}
\end{table*}

The percentage of large flares expected to be observed with the flare index strategy for a LEO mission ranges from $\approx$27\% for a mission that repoints weekly with a 24 hour response time (scenario F-1), to $\approx$38\% with a daily repoint at 24 hour response time (F-3), up to $\approx$42\% for a 12 hour daily response (F-4) and $\approx$47\% for a 1 hour daily response (F-5). Relying instead on the McIntosh classification yields similar but slightly inferior results, with $\approx$ 27\% of flares observed in the weekly repoint scenario (M-1), $\approx$37\% of flares observed at 24 hour daily response time (M-3), rising to 40\% at 12 hour daily response (M-4) and 41\% at 1 hour response (M-5). 
The Hale + number of spots strategy performs more poorly than both the flare index and McIntosh strategies, only managing to observe 36\% of flares with a 1 hour response time (H-5), a 10\% poorer performance than the flare index. This suggests that less useful information is contained in this target selection metric. As expected, choosing a target randomly is the worst approach, with a very poor mission performance. Using this approach, a mean of only $\approx$18\% of flares are observed, a value that is almost independent of response time. Since there is no skill associated with the target choice, improving the mission response time has no effect on performance. This strategy serves as a lower bound for the number of flares a partial field-of-view mission that targets active regions can expect to observe.

For the final scenario of each target method we consider a 1 hour response time and disregard eclipse and SAA intervals. This could be achieved for example by positioning the mission at L1. In this case, the performance distributions are very narrow as the only consideration is the response time, which in this case is drawn from a narrow distribution centered on 1 hour. This represents the best-case scenario, i.e. the maximum number of flares that a mission may expect to observe based on the chosen strategy. As expected, the random target selection still performs poorly, capturing $\approx$25\% of flares (R-6). The Hale strategy captures $\approx$48\% of events (H-6). The McIntosh and flare index strategies remain the best performing, observing $\approx$54\% (M-6) and $\approx$62\% (F-6) of flares respectively. Note that the flare observation percentages do not increase by the same factor as the improvement in the duty cycle. This is because a flare would have to be fully obscured by gaps in the duty cycle to be excluded from the previous scenarios. In reality, since flare duration is often tens of minutes, many flares are still partially observed and hence included in observation totals. 

Overall, using these target selection strategies a partial field-of-view solar flare mission in LEO with daily response capability may expect to observe between 35\% - 47\% of solar flares of class M1 or greater. In terms of raw flare numbers, 47\% corresponds to $\approx$ 277 flare observations over the mission period 2011 February 1 through 2014 December 31. Meanwhile, a mission in geosynchronous orbit or at L1 can expect to observe between $\approx$ 48\% and $\approx$ 62\% of flares that occur. In this case, 62\% corresponds to $\approx$ 365 flare observations.   

\subsubsection{Association With CMEs}

It is well-known that solar flares are closely associated with coronal mass ejections (CMEs) \citep[e.g.][]{2003SoPh..218..261A, 2005JGRA..11012S05Y, 2008ApJ...673.1174Y, 2018SoPh..293...60M} and thus coordinated observations are desirable. With our simulations we can explore the number of CME-associated flares that were observed, using a list of flare-CME associations compiled by \citet{2019AGUFMSH11D3389A} based on the comprehensive SOHO/LASCO CME catalog\footnote{The LASCO CME catalog is available at: \url{https://cdaw.gsfc.nasa.gov/CME_list/}.} \citep{2009EM&P..104..295G}. In the 2011 February 1 to 2014 December 31 time interval, there were a total of 302 solar flares of class $>$M1 out of 589 that were associated with CMEs of good quality. For each mission targeting strategy, we can determine the number of CME-associated flare observations. Note that this does not guarantee useful observations of the CME, as that depends on a variety of factors including the position of the flare on the solar disk and the propagation direction and velocity of the CME. Using the daily response with $t_r(12,2,10)$ as a baseline as shown in Figure \ref{performance_example} we find that when the McIntosh-based strategy is employed (scenario M-4), 18\% $\pm$ 1\% of the 589 available flares are observed \textbf{and} CME associated, a raw count of $\approx$ 105 events on average. For the Hale-based strategy (scenario H-4) the fraction is 14.5 \% $\pm$ 0.5 \%, and for the flare index strategy (scenario F-4) we find 18\% $\pm$ 1\%. Using the randomized pointing strategy (scenario R-4) performance is naturally much worse, but even in this case we find that 9\% $\pm$ 1\% of all available flares are observed and CME associated, around $\approx$ 50 events on average. 

\subsubsection{Historical Comparison with the TRACE Mission}

As noted in Section \ref{introduction}, a number of previous solar missions have observed the Sun with partial fields of view. It is useful to compare the results of our simulations with the actual performance of a historical mission. An analogy to the mission simulations presented here is the TRACE satellite \citep{1999SoPh..187..229H}, which operated from 1998 until 2010 in a Sun-synchronous orbit. TRACE was an EUV and UV imager with high spatial resolution and a FOV of 8.5 $\times$ 8.5 arcminutes, similar to our baseline simulation FOV of 9.8 $\times$ 9.8 arcminutes. 

\begin{figure}
    \centering
    \includegraphics[width=12cm]{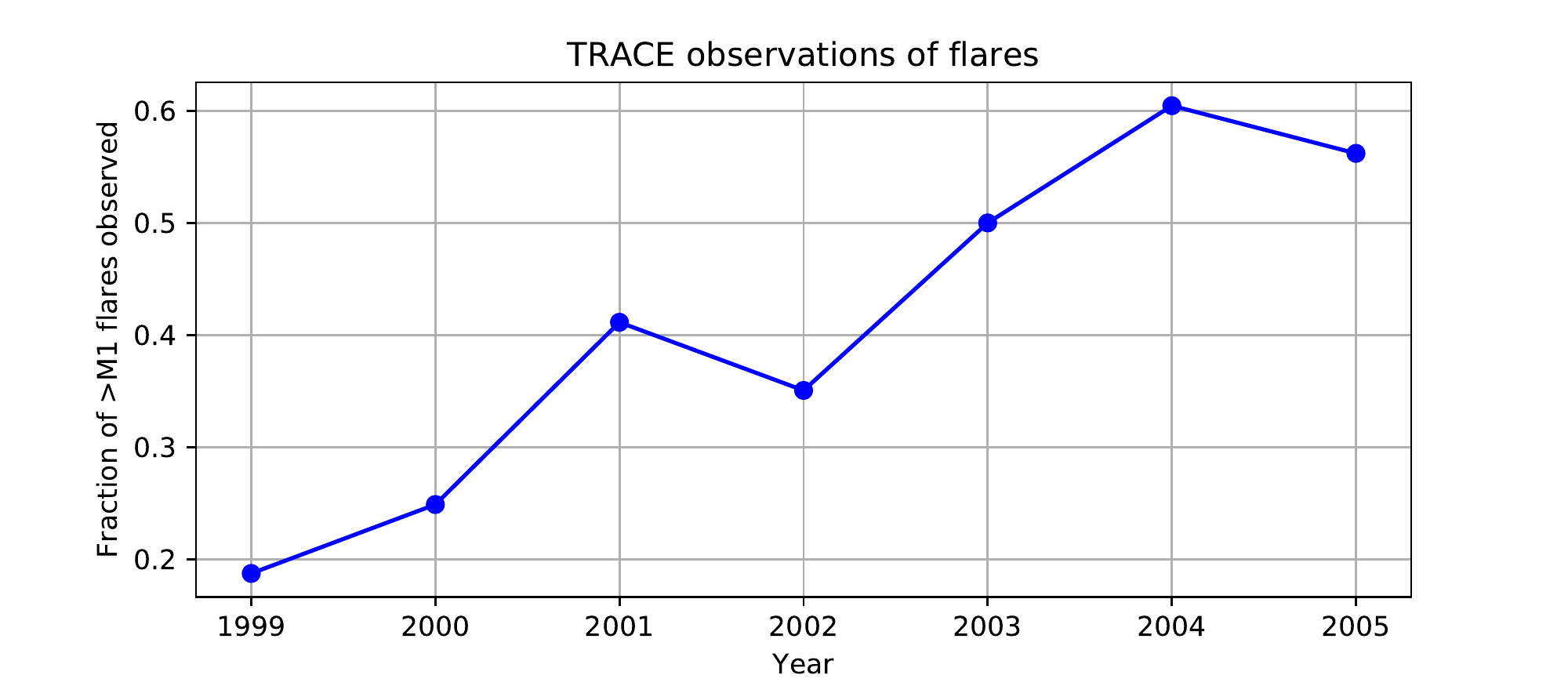}
    \caption{Fraction of all flares of GOES magnitude $>$ M1 self-reported as observed by TRACE each year between 1999 and 2005, according to the TRACE solar flare catalog. }
    \label{trace_performance}
\end{figure}

Figure \ref{trace_performance} shows the fraction of solar flares of magnitude $>$ M1 that TRACE observed each year between 1999 and 2006, according to the TRACE solar flare catalog\footnote{The TRACE flare catalog is available at: \url{http://helio.cfa.harvard.edu/trace/flare_catalog/}.}. Although mission operations continued until 2010, very few large flares occurred from 2006 onwards (see Figure \ref{solar_cycle}). Thus these years are excluded from analysis. The percentage of large flares observed by TRACE varies between 20\% in 1999 -- the first full year of operations -- to 60\% in 2004. In general, the fraction of flares observed by TRACE improves steadily over the course of Solar Cycle 23, possibly a reflection of changing science priorities over time.

This performance can be compared to our simulation results shown in Figure \ref{overall_performance}, reiterating the caveat that TRACE observed from a sun-synchronous orbit. We can see from Figure \ref{overall_performance} that a mission capable of daily repoints with a 1 hour delay and no eclipse or SAA intervals can expect to observe between 48\% and 62\% of large flares that occur on average, depending on the operations strategy. This is much better than the performance of TRACE early in Solar Cycle 23 between 1999 -- 2002, but broadly comparable to TRACE's performance during the latter portion the cycle, between 2003 -- 2005. 

\subsection{Active Region and Flare Data Exploration}
\label{ar_exploration}

The flare and active region data compiled here allows us to clarify some interesting aspects of flare distribution. Factors that directly impact how effective a partial field of view mission can be include how many active regions are typically present on the solar disk, how many flares occur per day, and how often multiple active regions produce flares concurrently.

\begin{figure}
    \centering
    \includegraphics[width=12cm]{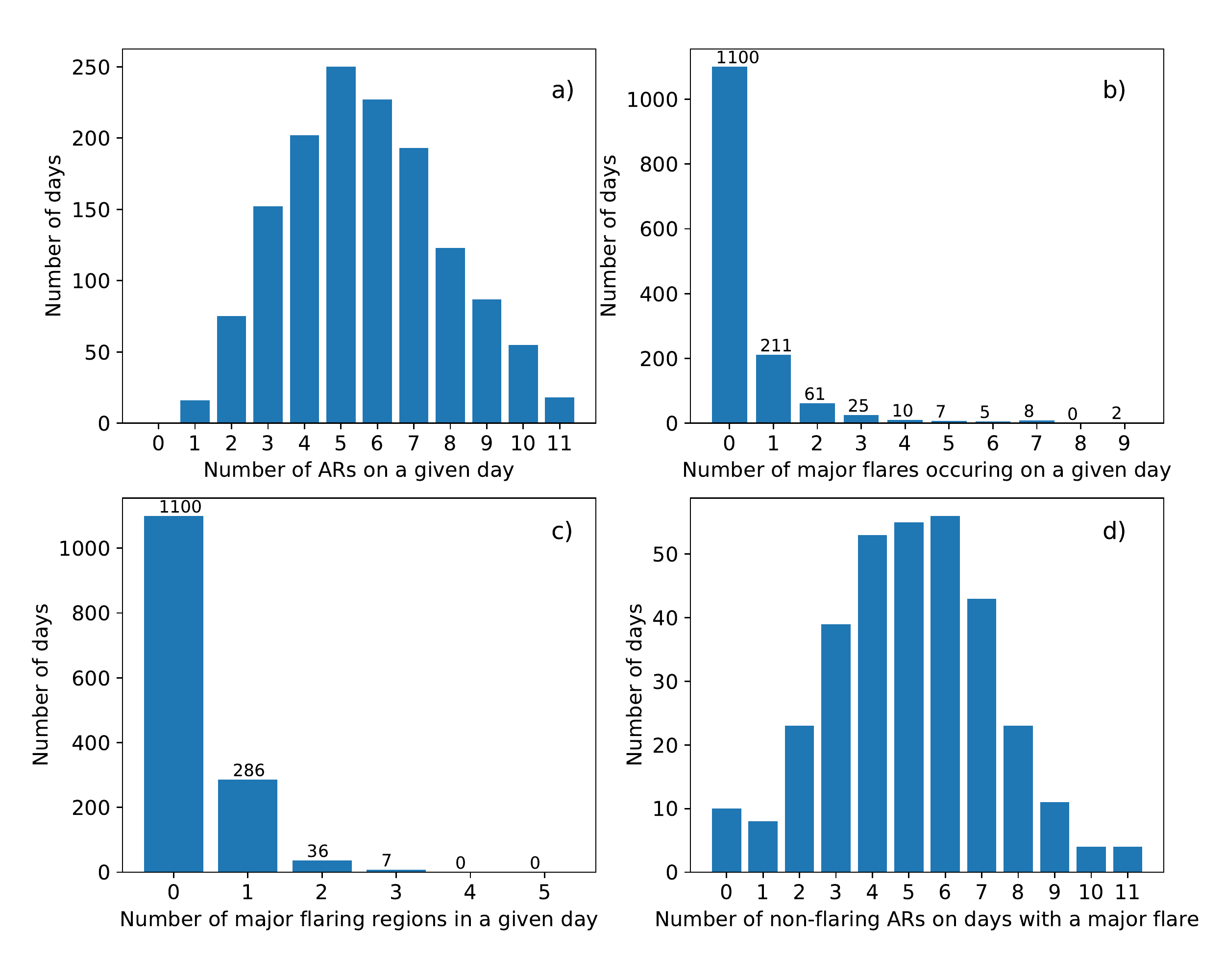}  
    \caption{Panel a): Total number of NOAA active regions present on the solar disk on a given day during the period 2011 February 1 -- 2014 December 31. Panel b): The number of flares of magnitude $>$ M1 that occurred on each day during the mission period. Panel c): The number of active regions producing major flares on a given day in the same period. Out of a total of 1429 days, 1100 of those days produced no flares of GOES class M1 or greater.  Panel d): The number of non-flaring ARs present on the Sun on days that produced at least one major flare.}
    \label{ar_figure}
\end{figure}

Figure \ref{ar_figure}a shows the number of NOAA-designated active regions present on a given day throughout the mission period. Clearly, there are almost always multiple ARs available to study, with a mean of $\approx$ 5 ARs present on a typical day. Figure \ref{ar_figure}b shows instead the number of solar flares $>$M1 that occurred on each mission day. Unlike the AR distribution, we can clearly see that most days (1100 out of 1429) produced no significant flares, with 211 days producing only one flare, and a total of 110 days that produced two or more. The most extreme values are two days during the mission that each produced 9 large flares. 

Since each flare in the GOES flare list is associated with a numbered NOAA active region, we can also determine how many different ARs are producing flares concurrently. For each day we determine the number of different active regions that produced major flares. The results are shown in Figure \ref{ar_figure}c, \ref{ar_figure}d.

Figure \ref{ar_figure}c reiterates that a large majority of days during the mission period, around 77\%, produced no flares of GOES class M1 or above. For the days that did produce major flares, they were most often sourced from a single NOAA active region (286 days or 20\% of total days), even though several other ARs were typically present on the disk (Figure \ref{ar_figure}d). In most of the remaining days, there were two active regions producing major flares (36 days, or 2.5\% of total days). Cases of more than two ARs producing major flares on the same day are rare, with only 7 occurrences throughout the mission. This illustrates the importance for a flare mission of choosing the correct active region target. Since only around 3\% of mission days involved two or active regions producing major flares, it follows that if flare prediction becomes sufficiently accurate and robust it will be theoretically possible to observe almost all major flares that occur on the Sun.

We can estimate this best possible outcome in a straightforward way. We simulate perfect predictive knowledge by assuming that on each day the mission pre-emptively chooses the target that would produce the greatest number of flares, and that the mission response time to new information is instantaneous, i.e. $t_r$ = 0. Furthermore, for simplicity we assume such a mission is at the L1 position such that eclipses and the South Atlantic Anomaly can be neglected. In this scenario, $\approx$ 90\% of all M and X-class solar flares can be observed during the mission. This provides a hypothetical reference point for the results presented in Section \ref{flare_numbers}. The remaining 10\% are missed because of days where multiple ARs are producing flares simultaneously (see Figure \ref{ar_figure}). Thus, improving beyond 90\% would require more frequent publication of input data products coupled with real-time mission response capabilities.

\subsection{Target Selection and Pointing Comparison}

\subsubsection{Mission Target Changes}

From the science planning perspective of a flare mission, a typical optimization is to maximize the number of flare observations while minimizing the number of targeting changes needed. Another important factor is the distribution of the chosen pointing locations on the Sun, and whether that distribution suffers from any bias.

\begin{figure}
    \centering
    \includegraphics[width=12cm]{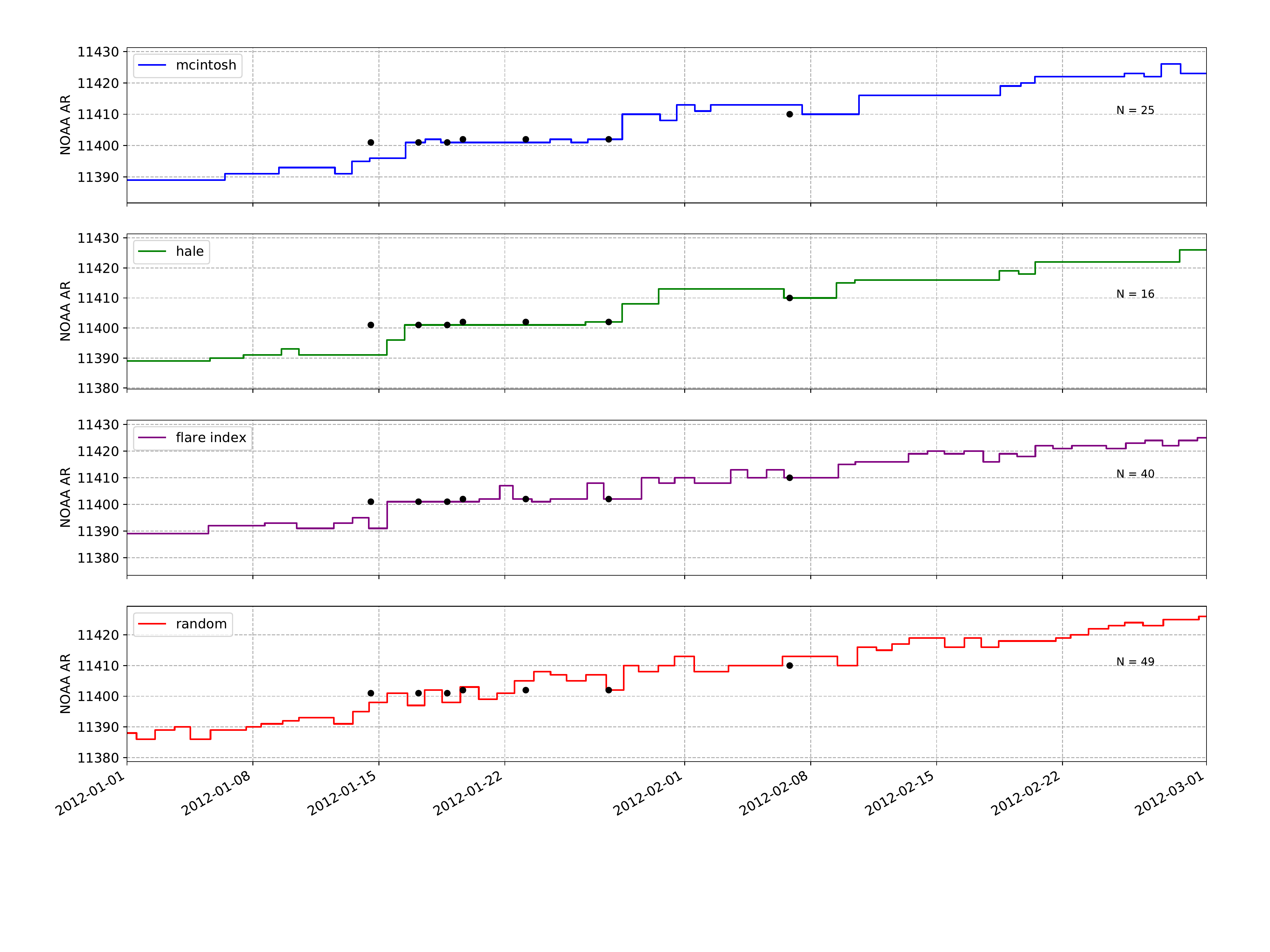}
    \caption{The mission AR target as a function of time between 2012-01-01 and 2012-03-01 for scenarios M-4 (1st row), H-4 (second row), F-4 (third row), and R-4 (fourth row). The y-axis shows the NOAA AR number that is considered to be the mission target at a given time. The individual dots represent 7 solar flares at their peak times matched to their source active regions. The $N$ values indicate the number of repoints that occur for each scenario.}
    \label{pointing_timeline}
\end{figure}

As we can see from Figure \ref{pointing_timeline}, each target determination method described in Section \ref{target_methods} results in a different pointing profile of the mission over time. This figure illustrates a timeline of the mission target for each method over a representative 2-month interval, between 2012 January 1 and 2012 March 1, with daily repoint capability a 12 hour response time, i.e. scenarios H-4, M-4, F-4 and R-4. We can see that, for scenario M-4 there are N\textsubscript{Mcintosh} = 25 pointing changes in the shown interval, and N\textsubscript{Hale} = 16 for scenario H-4. Meanwhile, using F-4 requires N\textsubscript{Flare} = 40 pointing changes over the same 2 month interval, while R-4 as expected requires the most overhead, with N\textsubscript{Random} = 49 target changes.

Over the full mission interval 2011 February 1 to 2014 December 31 we find that N\textsubscript{Mcintosh} = 543, N\textsubscript{Hale} = 520, N\textsubscript{Flare} = 770, and N\textsubscript{Random} = 1158. This corresponds to mean times between re-points of $\Delta$t\textsubscript{Mcintosh} = 2.6 days, $\Delta$t\textsubscript{Hale} = 2.8 days, $\Delta$t\textsubscript{Flare} = 1.9 days, and $\Delta$t\textsubscript{Random} = 1.2 days. The time between repoints for each scenario presented in Figure \ref{overall_performance} is shown in Table \ref{results_table}.

Of most interest may be the comparison between the McIntosh-based approach and the flare index approach. From Section \ref{flare_numbers} we determined that a target method based on recent flare activity demonstrated the best overall performance in terms of number of solar flares observed. From Figure \ref{pointing_timeline} we see that this comes at a cost in terms of the number of target changes required of the mission. In addition we also find that the Hale-based strategy not only under-performs the McIntosh-based approach significantly, but provides only a slight saving in terms of the number of mission pointings required.



\subsubsection{Spatial Distribution of Pointing Locations}

\begin{figure*}
    \centering
    \includegraphics[width=12cm]{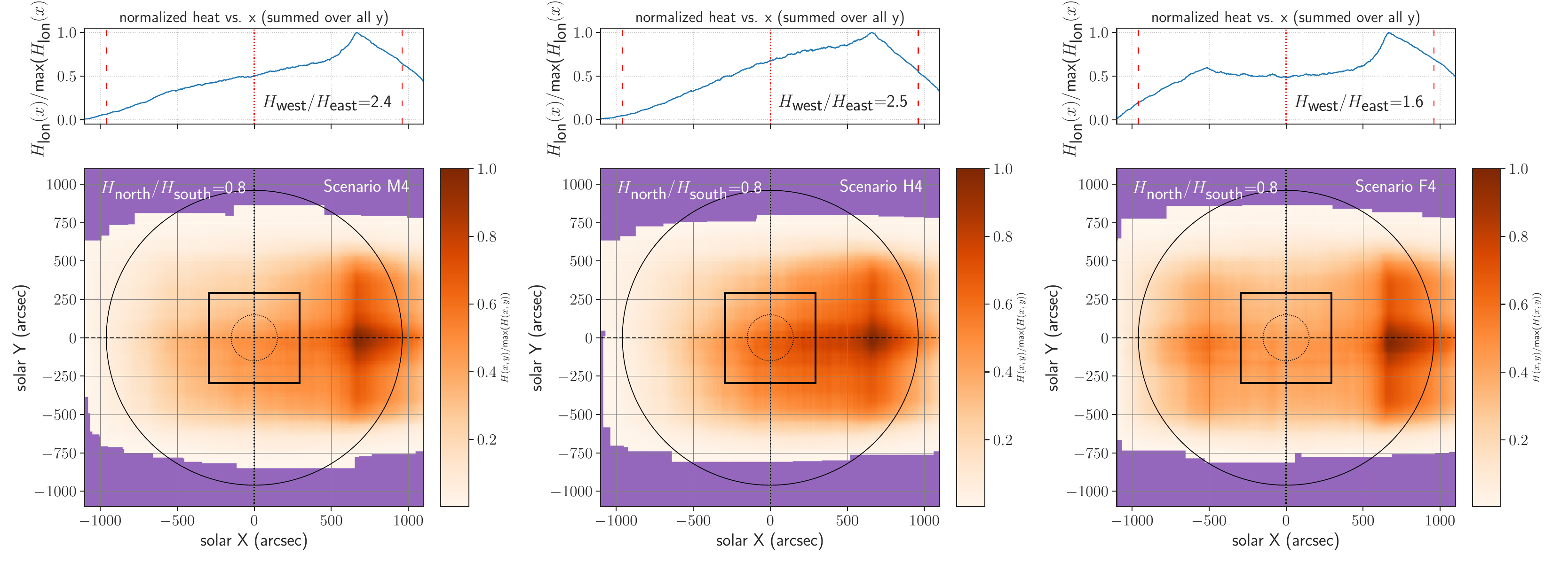}
    \caption{Heat maps of the mission pointing constructed by summing together the areas observed by the mission FOV over the duration of the mission. The maps shown are for Scenarios M-4 (left), H-4 (center) and F-4 (right). To build the heatmap, a snapshot of the FOV is extracted once per day between 2011 February 01 and 2014 December 31. The center map shows the full distribution of `heat' for the entire mission. The top panel shows the 1-d heat as a function of longitude. The purple shading highlights areas that are never in the field of view at any time. }
    \label{pointing_heatmap}
\end{figure*}

Also of interest is the spatial distribution of the mission field-of-view on the Sun. Figure \ref{pointing_heatmap} shows the daily distribution of the mission pointing for three scenarios; the McIntosh target method with a 12 hour mean repoint delay (M-4), the Hale target method with a 12 hour mean delay (H-4), and the flare index target method with a 12 hour mean delay (F-4). These distributions is obtained by summing and normalizing the number of times a given helioprojective $(x,y)$ coordinate appeared in the full mission field of view, and we refer to this as a `heat map'. We can see that for all scenarios the mission FOV is distributed relatively evenly between the northern and southern hemispheres, with a slight preference for the southern hemisphere. The ratio of north-to-south pointing heat is 0.8 in each case. This is consistent with the actual north-south distribution of solar flares during the mission period, which was $\approx$ 0.7. We can also compare this with the north-south distribution of source active regions, by extracting the latitudes of each active region reported every day of the simulation period in the SRS summaries. We find that the north-south distribution of active regions is $\approx$ 0.9. This differs from the flare ratio as it does not account for the complexity and flare productivity of the active regions.

\begin{figure}
    \centering
    \includegraphics[width=12cm]{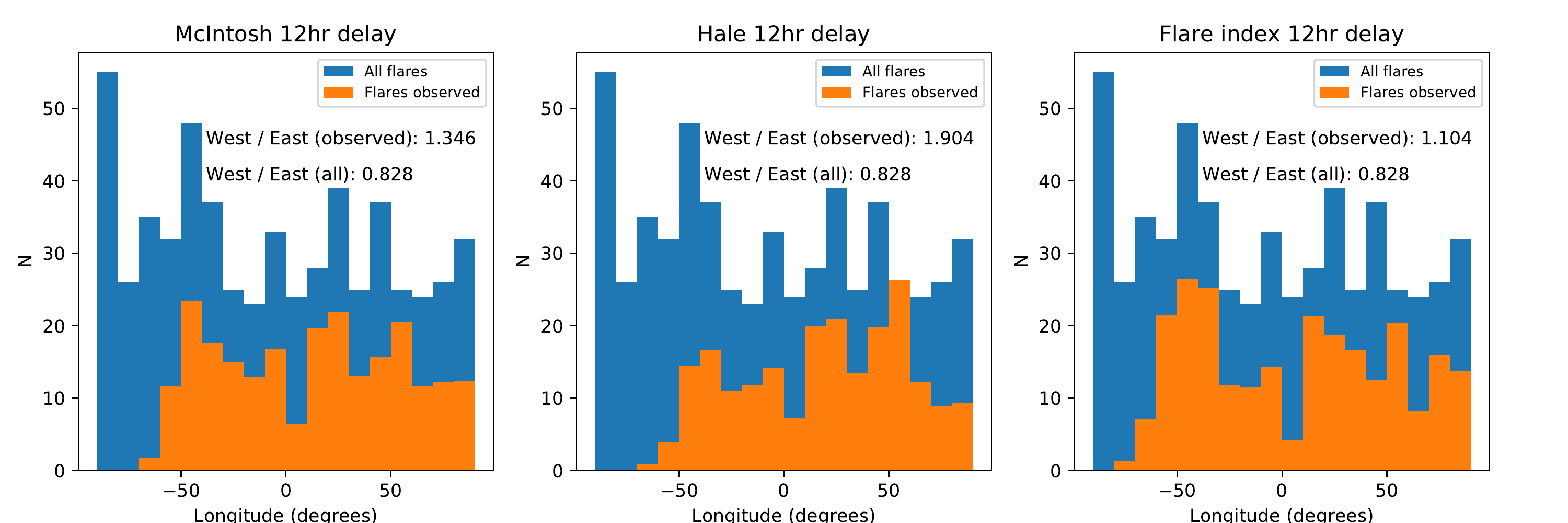}
    \caption{Flare heliographic longitude distributions for a) The McIntosh based target method, b) the Hale-based target method, and c) the flare index based target method. For each panel, the distribution of longitudes for every flare that occurred during the mission interval is shown by the blue histogram. For all flares, the West / East ratio is 0.83. The orange histograms show the longitude distribution of flares that were actually observed using the given target method, averaged over $N$=100 iterations.}
    \label{longitude_distribution}
\end{figure}

The distribution of pointing longitude however reveals a clear bias in favour of the western limb of the Sun. For the McIntosh scenario the west-to-east FOV ratio is $\approx$ 2.4, i.e. more than two-thirds of the FOV `heat' is located in the western solar hemisphere. This is despite the fact that the true distribution of flares during this period was slightly biased towards the eastern hemisphere (see Figure \ref{longitude_distribution}). Similarly, the west-to-east ratio for the Hale scenario is $\approx$ 2.5. The flare index scenario exhibits a smaller bias towards the west limb, with a west-to-east ratio of $\approx$ 1.6.

Figure \ref{longitude_distribution} illustrates further the bias in the mission pointing by showing the distribution of mission pointing locations for different target determination methods. For each panel the blue histograms show the distribution of flare longitudes for each flare that occurred during the mission period. The orange histograms instead show the longitude distributions of flares that were actually observed by the mission. For all flares, the parameter $N_{west}$ / $N_{east}$ = 0.83, indicating slightly more flares in the eastern hemisphere. However, for the McIntosh target method we find a West / East ratio of 1.35 for flares that were observed, and for the Hale-based method we find a ratio of 1.9, a substantial bias towards the western hemisphere in both cases. However, for the flare index based target method (Figure \ref{longitude_distribution}c) we find a West / East ratio of 1.1 for observed flares, much closer to the overall flare longitude distribution. All three methods however show a deficit in observations for events near the east limb, with longitudes further east than -70$^{\circ}$. These ratios are also less imbalanced than those found in the heatmaps shown in Figure \ref{pointing_heatmap}, indicating that some of the heatmap bias is generated during times where no major flares are occurring.

\begin{figure}
    \centering
    \includegraphics[width=12cm]{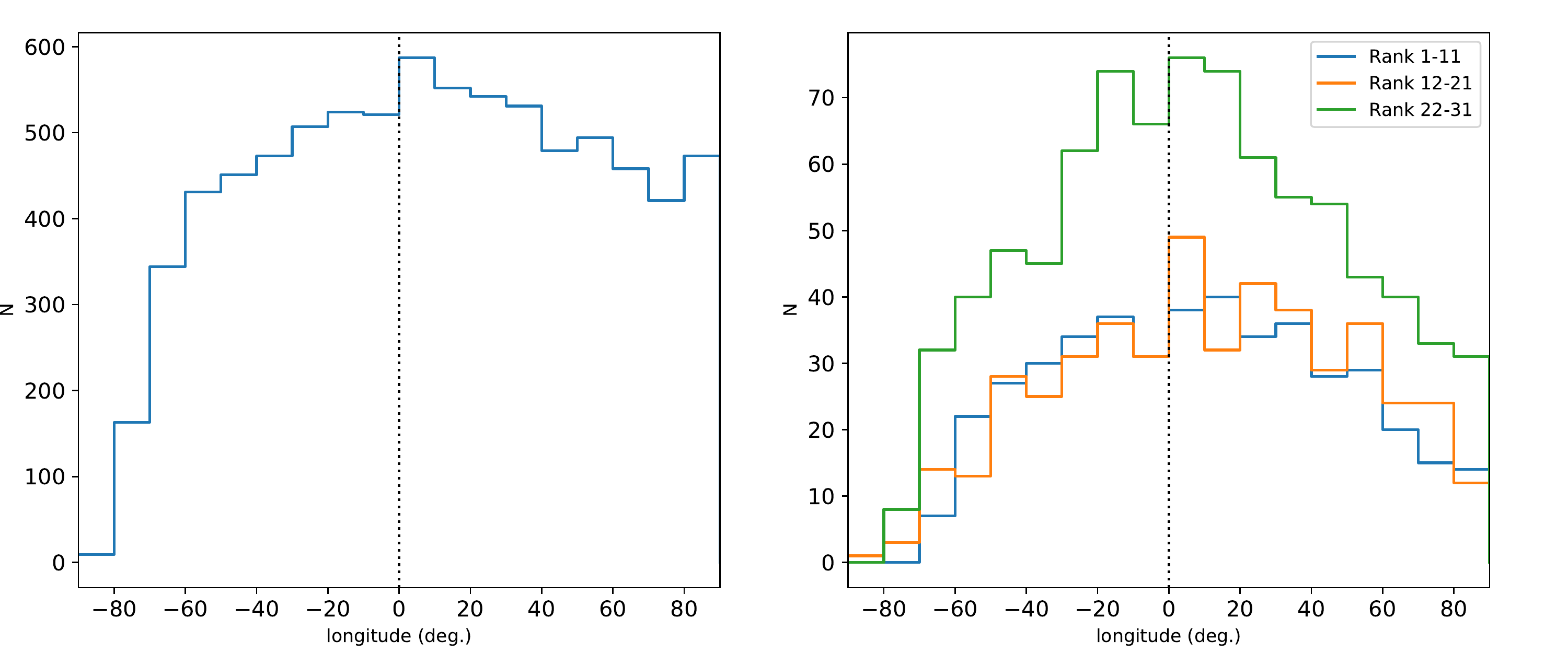}
    \caption{The distribution of different McIntosh active region classifications (as measured by their rank-ordered flare productivity) as a function of heliographic longitude throughout the mission period, derived from daily AR data. A deficit of classifications of all types is present towards the eastern limb of the Sun. Left: The distribution of all ARs with a Mcintosh classification as a function of longitude. Right: The same for Mcintosh classifications divided into three categories based on their rank-ordered flare productivity. Flare productivity is not linearly varying between ranks. Ranks 1 -- 11 represent the 11 most flare-productive classes that occurred during the mission period, while ranks 12 -- 21 represent the next most productive classes, and 22 -- 31 the next most productive after that. For clarity, ranks 32 and below are not shown in this panel.}
    \label{mcintosh_distributions}
\end{figure}

The bias of the mission pointing towards the western limb for certain target methods reveals a limitation in the available active region data, which is sourced from instrumentation along the Earth-Sun line. As new ARs rotate onto the visible Sun, there is a lag time before their properties can be confidently measured, due to the strong line-of-sight effects near the limb. Due to this, ARs are likely assigned to a lower-complexity category in the McIntosh or Hale classifications, or simply not categorized at all, until the AR rotates sufficiently onto the solar disk. This is shown in Figure \ref{mcintosh_distributions}, which was created by extracting the McIntosh classification and longitude for every NOAA-listed active region each day using the SRS data (see Section \ref{data}). In total, 53 different McIntosh classes occurred during the mission period, resulting in 42 unique ranks (since some classes have identical productivity values). Each McIntosh classification is then given a rank order based on its estimated flare productivity (see Section \ref{target_methods}), with rank 1 being the most flare-productive. Based on this, a distinct deficit of McIntosh classifications of all types can be seen at longitudes further East than -70$^{\circ}$. Conversely, as an AR approaches the western limb, the same effect occurs but less severely. The difference between limbs can be explained by the fact that ARs rotating towards the west limb are more likely to retain a categorization due to the difficulty of obtaining robust new measurements. Thus, any method that relies primarily on AR classification is likely to suffer from a significant bias towards the west limb, as shown in Figure \ref{longitude_distribution}. However, this is not necessarily problematic in practical terms, provided sufficient flares are observed overall. Furthermore, western hemisphere solar eruptive events are typically more geo-effective; depending on the particular science question it may be desirable to retain such a bias.

\subsubsection{Reducing West / East Bias}

The West / East bias of flare observations discussed in the previous section is driven by the fact that ARs near or over the East limb may have limited information of complexity or flare productivity, or simply may not be numbered at all (see Figure \ref{mcintosh_distributions}). Observational bias also exists on the West limb but the effect is not as pronounced.
Thus, the pointing strategies will preferentially point at ARs in the western hemisphere, even if the estimated flare productivities are very low.

\begin{figure}
    \centering
    \includegraphics[width=10cm]{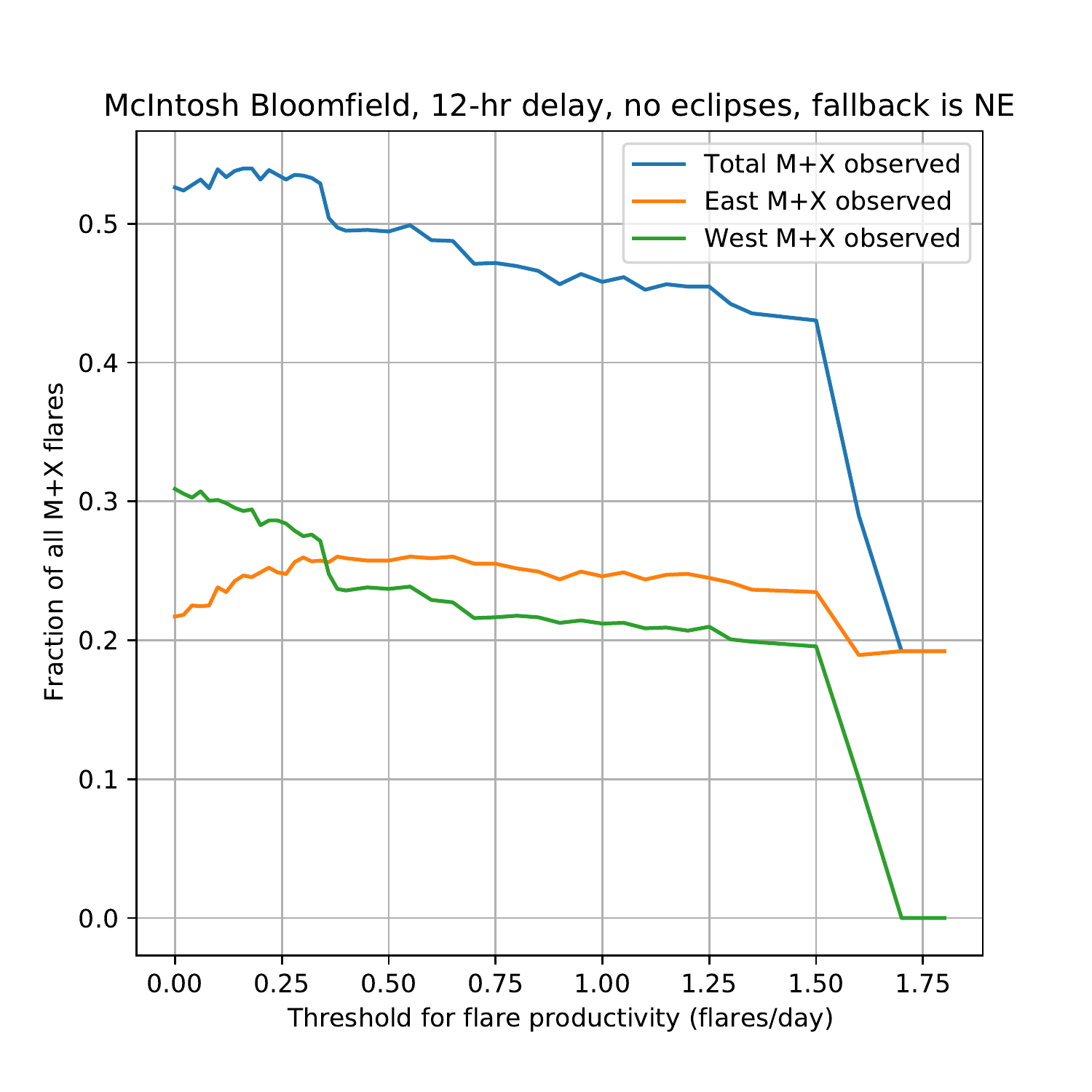}
    \caption{The observed fraction of $>$ M1 flares observed overall and from each hemisphere depends on the threshold for the minimum flare productivity of interest.  In this simulation, if no numbered ARs are above the threshold value, the observatory is pointed at ($-$650, +375) in helioprojective coordinates.  With zero threshold, the captured flares are heavily biased to the western hemisphere (green line much higher than orange line), but the West / East ratio becomes unity at a threshold of $\sim$0.36~flares/day. As the threshold increases beyond this value, the number of flares observed steadily decreases.}
    \label{threshold_scan}
\end{figure}

This bias can be reduced by modifying the relevant pointing strategies so that the observatory points near the East limb at active-region-band latitudes if the estimated flare productivities of the numbered ARs all fall below a threshold level.
Figure \ref{threshold_scan} shows an example of how, when the McIntosh target method is used, the West / East bias decreases as the threshold level is increased:

\begin{itemize}
    \item For low threshold levels ($\lesssim$0.33~flares/day), some western flares are missed, but additional eastern flares are observed, leading to a reduced West / East bias and even a small increase in the total number of flares observed.
    \item For moderate threshold levels ($\approx$0.5~flares/day), the number of observed eastern flares reaches a maximum, but the number of observed western flares is significantly impacted due to the ignoring of genuinely productive ARs.
    \item For higher threshold levels, the numbers of observed flares in each hemisphere and overall all steadily decline.
\end{itemize}

The appropriateness of such a threshold and the optimum threshold level will depend on the particular science question.

\subsection{The Effect of the Field-of-View}

In this Section, we investigate the importance of the mission field-of-view (FOV) in terms of maximizing the number of flares observed. For this purpose, we use simulations with McIntosh target method and a 12-hour mean repoint delay as our baseline (scenario M-4). Here, the mission is assumed to be in a 600 km LEO orbit and is thus affected by eclipses and SAA intervals. For each simulation run, we vary the FOV of the mission and measure the difference in the fraction of flares observed. 

\begin{figure}
    \centering
    \includegraphics[width=12cm]{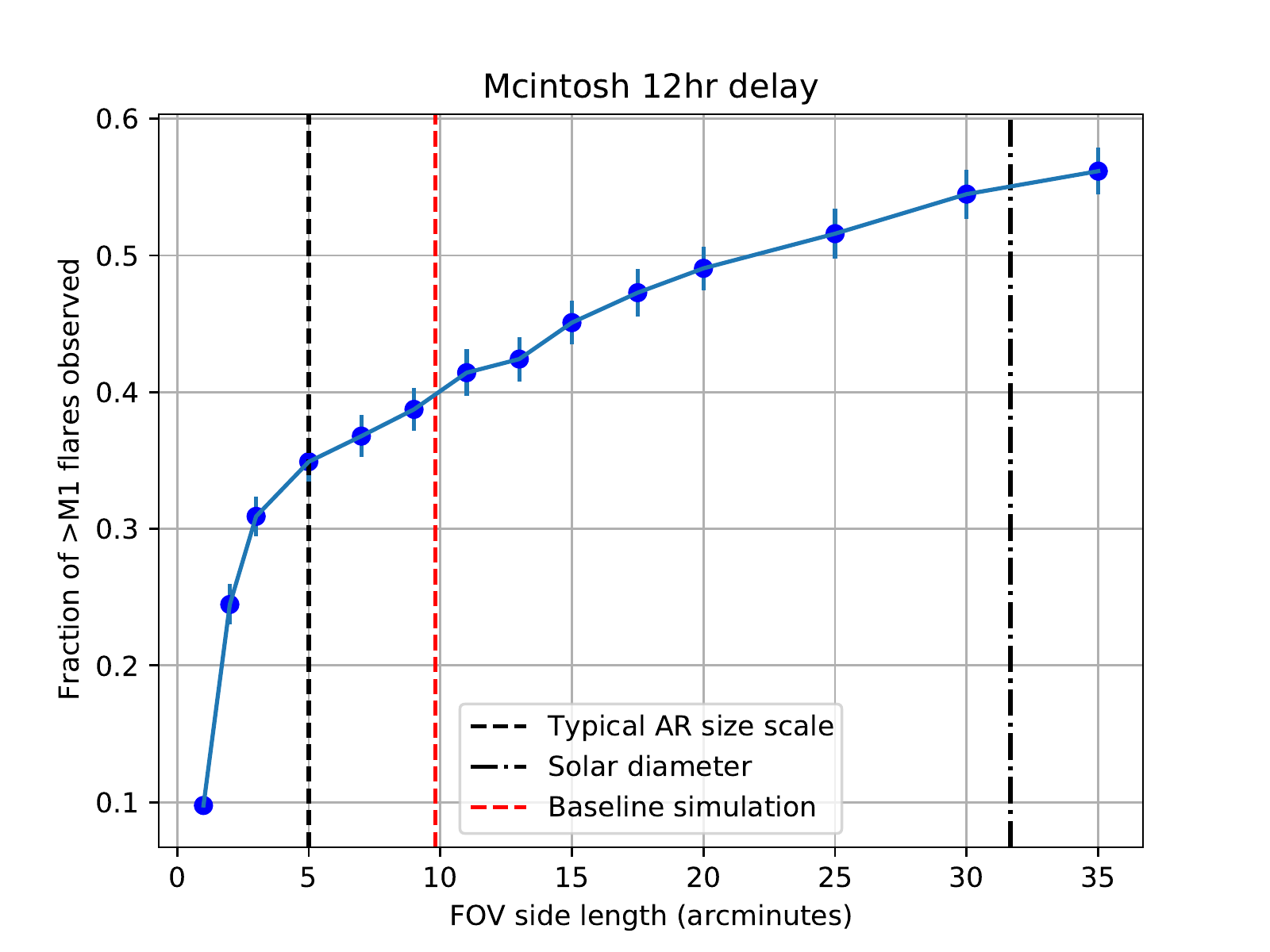}
    \caption{The fraction of solar flares of class $>$M1 observed during the mission period as a function of the FOV size. Here, to illustrate the effect of the FOV the McIntosh targeting method has been used for each simulation run, with a mean repoint delay of 12 hours (Scenario M-4). The error bar is the standard deviation of the distribution of fraction of flares observed.}
    \label{fov_figure}
\end{figure}

Figure \ref{fov_figure} shows the results of these simulations. Here the FOV is defined as a square with the side length given in arcminutes. In Figure \ref{fov_figure}a we observe a steep increase in the number of flares observed up to a FOV side length of $\approx$ 5 arcminutes. Above $\approx$ 5 arcminutes we observe a flattening, with the fraction of flares increasing approximately linearly from 5 to 25 arcminutes side length. These results are intuitive given typical active region sizes, which are typically less than 5 arcminutes in diameter \citep{1993SoPh..148...85H, 2017AN....338..398K}. Thus, once the FOV approaches and surpasses the active region size, we observe diminishing returns from increasing the FOV size further. This is due to the fact that there is often only one flaring active region on any given day (see Section \ref{ar_exploration}). The improvements in number of flares observed with large FOVs come from either the rare days where two or more active regions produce flares at the same time (a scenario that occurs on only 43 out of 1429 days -- or around 3\% of all days -- during the mission, as shown in Figure \ref{ar_figure}d), or days where the incorrect AR target was chosen but the FOV was sufficiently large to capture flares from a different target. Eventually, as the FOV encompasses the whole Sun, the choice of target becomes irrelevant and all flares are observed regardless, provided they are not obscured by eclipses or SAA intervals.

\section{Discussion}
\label{discussion}

We have performed simulations of the science operations and performance of a space-based, partial field of view solar flare mission. The key findings can be summarized as follows.

\begin{itemize}
    \item Based on daily input flare and active region data and a set of target selection methods, such a mission in low-earth orbit (with a duty cycle of 59\%) can expect to observe between 35\% and 47\% of large solar flares (M- and X-class) depending on the targeting strategy, how often the mission is able to re-point, and the time delay between receiving new information and re-pointing the spacecraft. In a control scenario, a mission operating in the absence of any flare-predictive information and choosing AR targets randomly should still observe $\approx$ 18\% of large flares. 
    \item In orbits with a 100\% solar observing duty cycle (e.g. a mission at L1) coupled with 1-hour re-point delays on average, between 48\% and 62\% of all large flares could be observed. 
    \item Of the target selection methods tested in this simulation, the overall best-performing approach was based on the flare index, a measure of flaring activity over the previous 24 hours. This approach yielded a slight improvement in number of flares observed compared to target selection based on McIntosh classifications, and a larger improvement over Hale classifications.
    \item A key factor determining the number of flares captured is the re-point delay $t_r$ between receiving new information and re-pointing the spacecraft. For example, for a mission choosing targets based on active region McIntosh classes will observe 37\% of large flares with a 24 hour mean delay time, compared to 42\% of large flares with a 1 hour mean delay time. Using the flare index target selection, 37\% of large flares were observed with a 24 hour mean delay time, rising to 46\% of large flares with a 1 hour mean delay time.
    \item The fraction of successful flare observations depends strongly on the chosen mission field-of-view, and falls off rapidly for FOV side lengths smaller than 5 arcminutes. At side lengths greater than 5 arcminutes, the FOV becomes larger than typical active region scales, thus gains in flare observation fractions are more gradual.
    \item Target selection methods based on active region complexities showed a significant pointing bias towards the western solar hemisphere. Using a McIntosh-based target selection method, the mission FOV was located in the western hemisphere approximately two-thirds of the time (Figure \ref{pointing_heatmap}). Target selection based on past flare activity instead resulted in approximately even distribution between hemispheres. This can be compensated for without reducing the number of flares observed by ignoring AR targets with low flare probabilities and instead targeting the east limb of the Sun.
    \item On any given day, major flares are most commonly sourced from a single active region. Thus, with sufficiently advanced flare prediction techniques, almost all major flares could be potentially observed even with a partial field of view mission.
\end{itemize}

These results provide a useful performance estimate for a future solar mission focused on flare observations, and inform the requirements that would ensure the success of such a mission. An L5 mission that observes the disk of the Sun at regular intervals would improve the East/West observation bias found in this study simply by providing information about active regions that will soon appear on the east limb as seen by a mission from the Earth/Sun line. Further, this study shows that a flare science mission along the Earth-Sun line can support missions at L5 without significant deleterious effects on the flare observation goals of the flare science mission.

The target selection methods presented in this work are straightforward and based on readily available current data products. An important future effort would be to compare these results with those obtained via simulations that incorporate human-based solar monitoring campaigns, such as the Max Millennium Major Flare Watch alerts\footnote{The Max Millennium program is described here: \url{http://solar.physics.montana.edu/max_millennium/}.}. A brief examination of the major flare watches issued during the simulation period indicates that the targets highlighted by the Max Millennium often, but not always, match those selected by the McIntosh and flare index target methods. Incorporating human-based alerts into future simulations may result in a modest improvement in the fraction of solar flares observed. We can compare our simulations with the overall performance of the Max Millennium Major Flare Watch program, which was studied by \citet{2016SoPh..291..411B} over the time period 2001 - 2010. \citet{2016SoPh..291..411B} showed that, if observers reacted immediately to each Major Flare Watch called, 56\% of X-class flares would have been caught and 45\% of $>$M5 flares would have been caught. Building in a 24 hr delay in responding decreased the X-flare efficiency only slightly to 53\%. These findings are comparable to the number of flares observed for our simulation runs of a mission with a 100\% duty cycle (e.g. Scenarios H-6, M-6, F-6). The efficiency of catching low M-class flares with the Major Flare Watch program was somewhat lower, but this can be explained by the fact that the program is designed specifically to capture strong flares of class M5 and greater.

A real future mission is likely to make use of multiple data sets simultaneously in order to make planning decisions, for example by combining active region complexities and recent flare activity. Moreover, the field of solar flare prediction is active and continually advancing, thus there is considerable scope for a future mission to improve over the performance estimated in this work.

\section*{Declarations of Potential Conflicts of Interest}
This work builds on analysis undertaken as part of the FOXSI SMEX Phase A study. The Phase A study was funded in part by funds from NASA and funds internal to NASA Goddard Space Flight Center. The authors declare that they have no conflicts of interest.

\begin{acks}
The authors would like to thank the anonymous referee, whose careful review improved the quality of this manuscript.
\end{acks}

\appendix




\section{Flare productivity of McIntosh classifications}
\label{mcintosh_appendix}

There are multiple estimates of the flare productivity of different McIntosh active region classifications. Many of these flare rates are based on relatively small number statistics and are subject to significant uncertainty. In the main body of the paper, we adopt the M-class flare rates published by \citet{2012ApJ...747L..41B}. However, it is useful to compare the results of the simulation using alternative sources. Combined M- and X-class flare rates were previously published by \citet{1994SoPh..150..127B}. Additionally, in \citet{1994SoPh..150..127B} it was suggested that flare productivity of any active region could be estimated by averaging the overall productivity of each of the three Mcintosh parameter values. One motivation for this approach is the ability to estimate the productivity of a region that has not been observed or has very few observations.

\begin{figure}
    \centering
    \includegraphics[width=10cm]{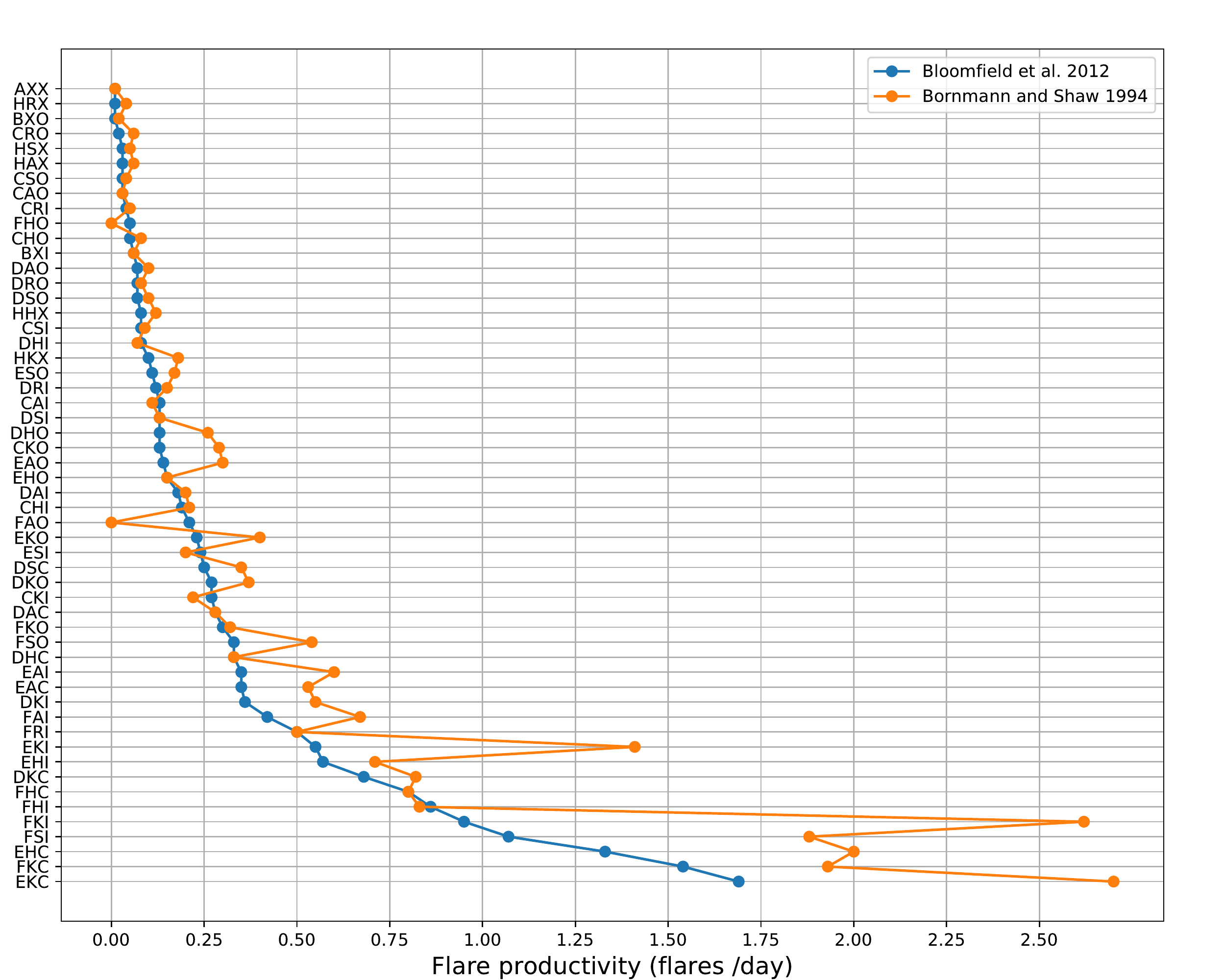}
    \includegraphics[width=10cm]{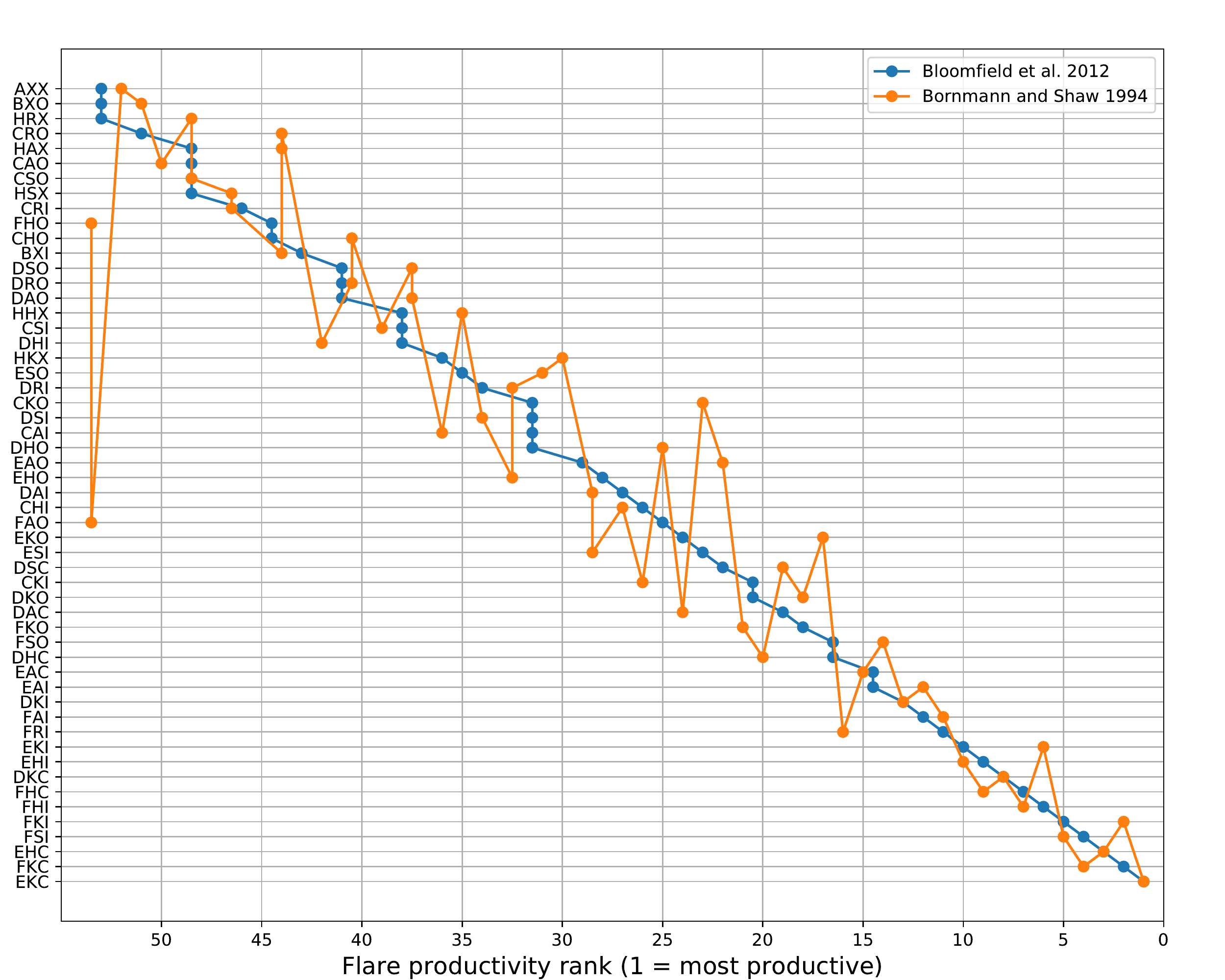}
    \caption{Top: The flare productivity estimates for each Mcintosh active region classification according to \citet{2012ApJ...747L..41B} (blue) and \citet{1994SoPh..150..127B} (orange). For the \citet{2012ApJ...747L..41B} estimates we take the values calculated for M-class solar flares, while the \citet{1994SoPh..150..127B} values combine M- and X-class solar flares. Bottom: The rank order of each of the Mcintosh active region classifications in terms of their flare productivity. Rank 1 is the most flare-productive classification. }
    \label{bloomfield_bornmann_comparison}
\end{figure}

Figure \ref{bloomfield_bornmann_comparison}a shows the flare productivity values published by \citet{2012ApJ...747L..41B} (blue) compared with the estimates by \citet{1994SoPh..150..127B} (orange). We can see that the overall trends in these datasets are consistent, with a few prominent outliers of note. Both datasets agree that the five most productive regions are those designated EKC, FKC, EHC, FSI, and FKI, with differences in the exact productivity values and ordering.

Figure \ref{bloomfield_bornmann_comparison}b illustrates the comparison of these datasets via their rank ordering, where rank 1 is the most flare-productive region. This reinforces that both datasets agree on the overall trend of productivities, with localised differences in the ordering. The most notable deviation is the classification FAO; in the original \citet{1994SoPh..150..127B} this classification was assigned a productivity of 0.0, however in the \citet{2012ApJ...747L..41B} it has an estimated productivity of 0.21 flares/day, the 25th most productive classification. The FHO classification was also assigned a productivity of 0.0 by \citet{1994SoPh..150..127B} but has a productivity of 0.05 in \citet{2012ApJ...747L..41B}. The other most notable discrepancy is that the FKI and EKI McIntosh classifications are both considered much more productive in the \citet{1994SoPh..150..127B} dataset compared with the \citet{2012ApJ...747L..41B} dataset, both in flare productivity value and in their rank order. 

\begin{figure}
    \centering
    \includegraphics[width=12cm]{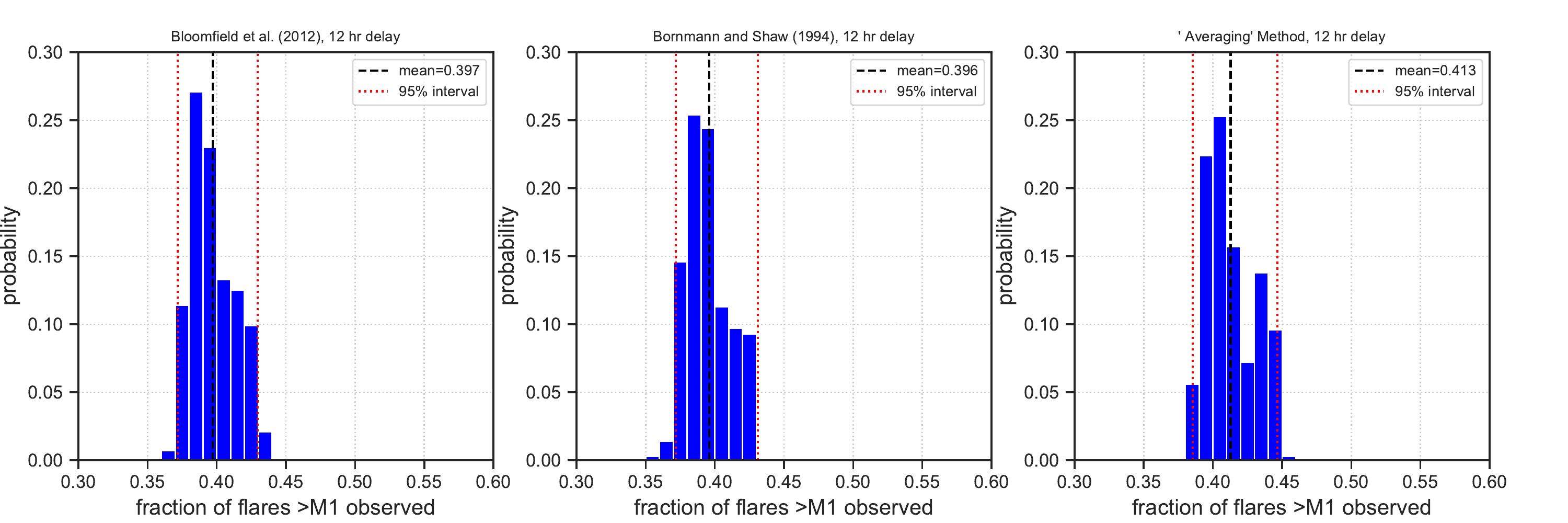}
    \caption{The flare observing performance of the mission using the McIntosh target method, using three different ways of estimating the flare productivity of Mcintosh classifications. Left panel: Performance of the mission using flare productivities calculated using the \citet{2012ApJ...747L..41B} M-class flare productivities. This is used in the main body of the paper. Center: Mission performance using the flare productivity estimates from \citet{1994SoPh..150..127B}. Right panel: Mission performance obtained by averaging the overall productivity of each individual Mcintosh parameter, as originally suggested by \citet{1994SoPh..150..127B}. }
    \label{bloomfield_bornmann_histogram_comparison}
\end{figure}

Figure \ref{bloomfield_bornmann_histogram_comparison} shows the impact of adopting these different methods of ranking Mcintosh classifications on the mission performance. This simulation used the 12-hour delay scenario (see Section \ref{flare_numbers}). From this it is clear that the difference is small; using the \citet{2012ApJ...747L..41B} flare productivities a mean of 39.7\% of $>$M1 flares was observed, whereas when the \citet{1994SoPh..150..127B} data were adopted the mean percentage of flares observed was 39.6\%. Interestingly, the 'averaging' method proposed by \citet{1994SoPh..150..127B} performs slightly better than the explicit data-based approaches, observing a mean of 41.3\% of available flares. However, this approach is more challenging to scientifically justify, and we therefore choose not to adopt it in the main body of this work.

\bibliographystyle{spr-mp-sola}
\bibliography{refs.bib}

\end{article}

\end{document}